\documentclass[11pt]{article}

\date{\today}

\usepackage{amsmath,amssymb,amsthm}
\usepackage{color}
\usepackage[thinlines,thiklines]{easybmat}

\usepackage{algorithm}
\usepackage{algorithmic}
\usepackage{enumitem}

\usepackage{verbatim}
\usepackage{palatino}
\usepackage{mathpazo}
\usepackage{stmaryrd}
\usepackage[margin=1in]{geometry}
\usepackage{mathtools}
\mathtoolsset{centercolon}

\usepackage{graphicx}
\usepackage{amsfonts}

\def\bra#1{{\langle#1|}}
\def\ket#1{{|#1\rangle}}

\def\proj{{\hat{\cal P}}}

\def\H{{\hat H}}

\def\id{{\hat I}}
\def\bra#1{{\langle#1|}}
\def\ket#1{{|#1\rangle}}

\def\proj{{\hat{\cal P}}}

\def\H{{\hat H}}

\def\proj{{\hat{\cal P}}}

\def\id{{\hat I}}
\def\cB{{\cal B}}
\def\cC{{\cal C}}
\def\cD{{\cal D}}
\def\cE{{\cal E}}
\def\cG{{\cal G}}
\def\cH{{\cal H}}
\def\cS{{\cal S}}
\def\cV{{\cal V}}

\def\bI{{\mathbf I}}
\def\ba{{\mathbf a}}
\def\bb{{\mathbf b}}
\def\bee{{\mathbf e}}
\def\br{{\mathbf r}}
\def\bu{{\mathbf u}}
\def\bw{{\mathbf w}}
\def\bal{{\mathbf\alpha}}
\def\bbe{{\mathbf\beta}}
\def\bo{{\mathbf 0}}
\def\bbZ{{\mathbb{Z}}}

\def\T{{\tt{T}}}

\title{Entanglement-Assisted Quantum Error-Correcting Codes}

\author{Todd Brun $^1$ and Min-Hsiu Hsieh $^2$
\\[4mm]
\textit{$^1$ University of Southern California, USA}\\ 
\textit{$^2$ Centre for Quantum Software and Information, University of Technology Sydney, Australia}}

\date{}

\begin{document}

\pagenumbering{roman}
\maketitle
\tableofcontents
\cleardoublepage
\pagenumbering{arabic}


\section{Introduction}

Quantum error-correcting codes (QECCs) have turned out to have many applications; but their primary purpose, of course, is to protect quantum information from noise.  This need manifests itself in two rather different situations.  The first is in quantum computing.  The bits in a quantum computer are subject to noise due both to imprecision in the operations (or gates) and to interactions with the external environment.  They undergo errors when they are acted upon, and when they are just being stored (though hopefully not at the same rate).  All the qubits in the computer must be kept error-free long enough for the computation to be completed.  This is the principle of fault-tolerance.

A rather different situation occurs in quantum {\it communication}.  Here, the sender and receiver (Alice and Bob) are assumed to be physically separated, and the qubits travel from the sender to the receiver over a (presumably noisy) channel.  This channel is assumed to be the dominant source of errors.  While the qubits may undergo processing before and after transmission, errors during this processing are considered negligible compared to the channel errors.  This picture of transmission through a channel is very close to the picture underlying classical information theory.

The types of codes that are of interest in quantum computation and quantum communication may be quite different.  In computation, reliability is key.  Therefore, codes with a high distance (such as concatenated codes) are used.  Also, the choice of code may be highly constrained in other ways:  for instance, it is very important that the codes allow efficient circuits for encoded logic gates (such as the transversal gates allowed by the Steane code).  The more efficient these circuits, the better the fault-tolerant threshold.  (See \cite[chapter 5]{LB13}  for a fuller discussion of this.)

By contrast, in communication one often imagines sending a very large number of quantum bits, perhaps even a continuous stream of qubits (as in the convolutional codes of \cite[chapter 9]{LB13}).  Here, one is generally trying to maximize the rate of transmission, often by encoding many qubits into a large block, within the constraint of a low error probability for the block.  In the asymptotic limit, one would like to achieve the actual capacity of the channel---the maximum rate of communication possible.

Because the sender and receiver are separated, joint unitary transformations are impossible.  However, they might be able to draw on other resources, such as extra classical communication, pre-shared randomness, or pre-shared entanglement.  In this chapter we study the use of pre-shared entanglement in quantum communication, and how we can design QECCs that use entanglement to boost either the rate of communication or the number of errors that can be corrected.  Codes that use pre-shared entanglement are called {\it entanglement-assisted quantum error-correcting codes} (EAQECCs).  We will study a large class of these codes that generalizes the usual stabilizer formalism from \cite[chapter 2]{LB13}.  We will also see how these codes can readily be constructed from classical linear codes, and how they can be useful tools in building standard QECCs.

\subsection{Entanglement-assisted codes}

In a standard QECC, the encoding operation proceeds in two steps.  First, to the quantum state $\ket\psi$ of $k$ qubits that one wishes to encode, one appends some number of ancilla qubits in a standard state (usually $\ket0$), and then applies an encoding unitary $U_{\rm enc}$:
\begin{equation}
\ket\psi \rightarrow \ket\psi \otimes \ket0^{\otimes n-k}
\rightarrow \ket{\Psi_L} = U_{\rm enc} \ket\psi \otimes \ket0^{\otimes n-k} ,
\end{equation}
where $\ket{\Psi_L}$ is the encoded or logical state.  The systems in states $\ket{\psi}$ and $\ket0$ are initially in the possession of Alice, the sender.  The encoding unitary acts on the space of the input qubits and ancillas together.

In an EAQECC, one can append not only ancillas, but also ebits, before doing the encoding unitary:
\begin{equation}
\ket\psi \rightarrow \ket\psi \otimes \ket0^{\otimes n-k-c} \otimes \ket{\Phi_+}_{AB}^{\otimes c}
\rightarrow (U_{\rm enc}\otimes\id_B) \ket\psi \otimes \ket0^{\otimes n-k} \otimes \ket{\Phi_+}_{AB}^{\otimes c} .
\end{equation}
The states $\ket{\Phi_+}_{AB}$ are EPR pairs shared between the sender (Alice) and the receiver (Bob).  The encoding operation $U_{\rm enc}$ acts on the qubits in Alice's possession; we write $(U_{\rm enc}\otimes\id_B)$ above to indicate that the encoding acts as the identity on Bob's halves of the ebits.  Obviously, in order to append $c$ ebits to the information qubits, Alice and Bob must have $c$ ebits of pre-shared entanglement.  After Alice does the encoding, all of her qubits are sent through the channel, including the $c$ halves of ebits.  So this procedure consumes $c$ ebits of preshared entanglement.

\begin{figure}[htbp]
\begin{center}
\includegraphics[width=4in]{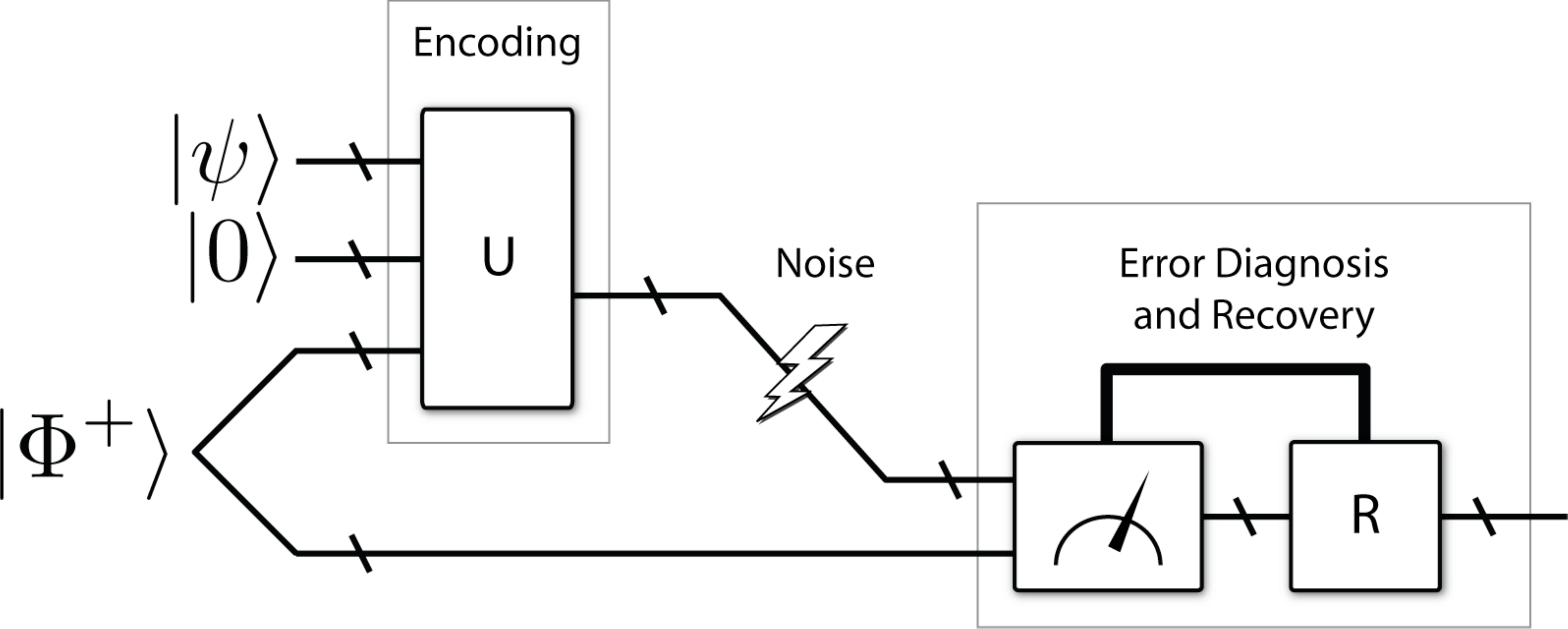}
\caption{Schematic structure of an entanglement-assisted quantum error-correcting code.  Alice and Bob share ebits that Alice uses in encoding her quantum information; these qubits all pass through the noisy channel, but Bob's halves of the initial ebits do not.}
\label{eaqecc_schematic}
\end{center}
\end{figure}

However, note that Bob's halves of the $c$ ebits do not pass through the channel.  These ebits were prepared ahead of time, by entanglement distillation or a similar procedure (see \cite{BBPSSW96,BDSW96} and \cite[chapter 22]{LB13}).  Because Bob's qubits do not pass through the channel, they are assumed to be error-free.

Why should we wish to use shared entanglement in our encoding?  There are two ways to see how this could enhance the power of the code.  First, we can compare EAQECCs to {\it teleportation}.  In teleportation, an ebit can be used (in combination with classical communication) to transmit one qubit perfectly from the sender to the receiver.  We might therefore expect that making use of ebits in error-correction could boost the rate of transmission.

Second, we can compare EAQECCs to {\it superdense coding}.  In superdense coding, by using an ebit the sender can send two classical bits of information to the receiver by means of a single qubit.  Each ancilla bit in a standard QECC can be thought of as holding one classical bit of information about any errors that have occurred.  Replacing an ancilla with one half of an ebit could, in principle, allow the receiver to extract {\it two} bits of classical information about the errors; with more information, more errors can be corrected.  We shall see, in the constructions that follow, that we can think of the enhanced power of EAQECCs in both of these ways, but the superdense coding interpretation is more often the best fit.  EAQECCs can achieve some communication tasks with fewer resources (or less chance of error) than would be needed by a standard QECC plus teleportation \cite{HW10}.

The possibility of constructing entanglement-assisted codes was suggested in \cite{BDSW96}, though no practical codes were constructed.  From asymptotic results in quantum information theory, we know that having pre-shared entanglement allows an enhanced rate of quantum communication between a sender and receiver \cite{DHW04,DHW08}.  EAQECCs are the finite-length realization of this idea.  By constructing larger and larger code blocks, in principle one can approach the entanglement-assisted quantum capacity of the channel.  An example of such a code was constructed by Bowen in \cite{Bow02}, starting from a standard stabilizer QECC (see \cite[chapter 2]{LB13}).  In \cite{BrunDevetakHsieh06, BDH06b, HDB07} an essentially complete theory of stabilizer EAQECCs was constructed.  That is the theory presented in this chapter.

Note, however, that in general shared entanglement does not come for free: entanglement must be established between the sender and receiver, and this in general requires the use of quantum channels, and perhaps additional entanglement purification.  Therefore, EAQECCs do not outperform standard quantum codes under all circumstances.  The enhanced communication rates that come from the use of entanglement must be paid for in establishing the entanglement in the first place.  For many purposes, what is important is the {\it net rate} of the code:  the number of qubits transmitted minus the number of ebits consumed.  With this measure, EAQECCs tend to perform about the same as standard codes.  Even in this case, though, EAQECCs sometimes have advantages:  as we shall see, there are many fewer restrictions on the construction of EAQECCs than standard codes.

\section{Constructing EAQECCs}
\label{constructingEAQECCs}

\subsection{Noncommuting ``stabilizers''}

First, let us recall the stabilizer formalism for conventional
quantum error-correcting codes, as presented in \cite[chapter 2]{LB13}.
Let $G_n$ be the $n$-qubit Pauli
group \cite{NC00}.  Every operator in $G_n$ either has eigenvalues
$\pm1$ or $\pm i$. Let $S\subset G_n$ be an Abelian subgroup
that does not contain $-I$.  Then this subgroup has a common
eigenspace $C(S)$ of $+1$ eigenvectors, which we call the {\it
code space} determined by the stabilizer $S$. Later on, we will
just use $\cC$ to denote the code space. Typically, the stabilizer is
represented by a minimal generating set $\{g_1,\ldots,g_m\}$, which
makes this a compact way to specify a code (analogous to
specifying a classical linear code by its parity-check matrix).  We
write $S = \langle g_1,\ldots,g_m\rangle$ to denote that $S$ is
generated by $\{g_1,\ldots,g_m\}$.

Let $\cE\subset\cG_n$ be a set of possible errors. If a particular
error $E_1\in \cE$ anticommutes with any of the generators of $S$,
then the action of that error can be detected by measuring the
generators; if the measurement returns $-1$ instead of $1$, we know
an error has occurred.  On the other hand, if the error is
actually {\it in} the stabilizer $S$, then it leaves all the
states in $\cC$ unchanged.  The code
$\cC$ can correct any error in $\cE$ if either $E_2^\dagger E_1 \notin
Z(S)$ or $E_2^\dagger E_1 \in S$ for all pairs of errors $E_1$ and $E_2$ in
$\cE$, where $Z(S)$ is the {\it centralizer} of $S$.

We can now generalize this description to the entanglement-assisted
case.  Given a \emph{nonabelian} subgroup ${\cal S}\subset{\cal
G}_n$ of size $2^m$, there  exists a set of generators
$\{\bar{Z}_1,\cdots,\bar{Z}_{s+c},\bar{X}_{1},\cdots,\bar{X}_{c}\}$
for $\cal S$ with the following commutation relations:
%
\begin{eqnarray}
{}[\bar{Z}_i,\bar{Z}_j] = 0 && \forall {i,j}  \nonumber\\
{}[\bar{X}_i,\bar{X}_j] = 0 && \forall {i,j}  \nonumber\\
{}[\bar{X}_i,\bar{Z}_j] = 0 && \forall {i\neq j} \nonumber\\
{}\{\bar{X}_i,\bar{Z}_i\} = 0 && \forall {i}.
\label{comm}
\end{eqnarray}
(We will see shortly how to find this set of generators for any subgroup.)
Here, $[A,B]$ is the commutator and $\{A,B\}$
the anti-commutator of $A$ with $B$.
The parameters $s$ and $c$ satisfy $s+2c=m$. Let $S_I$ be the
\emph{isotropic} subgroup generated by
$\{\bar{Z}_{c+1},\cdots,\bar{Z}_{c+s}\}$ and $S_E$ be the
\emph{entanglement} subgroup generated by
$\{\bar{Z}_{1},\cdots,\bar{Z}_{c},\bar{X}_{1},\cdots,\bar{X}_{c}\}$.
The numbers of generators of $S_I$ and pairs of generators of $S_E$
describe the number of ancillas and
the number of ebits, respectively, needed by the corresponding EAQECC.
The pair of subgroups $(S_I,S_E)$ defines an $[[n,k;c]]$ EAQECC
$\cC(S)$.  We use the notation $[[n,k;c]]$ to denote an EAQECC that
encodes $k=n-s-c$ logical qubits into $n$ physical qubits, with the help of
$s$ ancillas, and $c$ ebits  shared between
sender and receiver. These $n$ qubits are transmitted from
Alice (the sender) to Bob (the receiver), who measures them together
with his half of the $c$ ebits to correct any errors and
decode the $k$ logical qubits. We define $k/n$ as the {\it rate} and
$(k-c)/n$ as the {\it net rate} of the code.
(Sometimes we will write $[[n,k,d;c]]$ to indicate that the ``distance'' of the
code is $d$, meaning it can correct at least $\lfloor\frac{d-1}{2}\rfloor$
single-qubit errors.)

\subsection{The Gram-Schmidt Procedure}

We now illustrate the idea of the entanglement-assisted stabilizer
formalism by an example.  Let $\cS$ be the
group generated by the following non-commuting set of operators:
\begin{equation}
\begin{array}{ccccc}
\label{setM}
M_1= & Z & X & Z & I   \\
M_2= & Z & Z & I & Z   \\
M_3= & X & Y & X & I   \\
M_4= & X & X & I & X
\end{array} .
\end{equation}
It is easy to check the commutation relations of this set of
generators: $M_1$ anti-commutes with the other three generators,
$M_2$ commutes with $M_3$ and anti-commutes with $M_4$, and $M_3$
and $M_4$ anti-commute.  We will begin by finding a different set of
generators for $S$ with a particular class of commutation
relations.  We then relate $S$ to a group $B$ with a
particularly simple form, and discuss the error-correcting
conditions using $B$.  Finally, we relate these results back to
the group $S$.

To see how this works, we need two lemmas.  (We only sketch the proofs---these
results are well known.)  The first lemma shows
that there exists a new set of generators for $S$ such that $S$
can be decomposed into an ``isotropic'' subgroup $S_I$, generated by a set of
commuting generators, and a ``symplectic'' subgroup $S_S$, generated by a set of
anti-commuting generator pairs \cite{FCY04}.

\medskip

\noindent {\it Lemma 1.} Given any arbitrary subgroup $V$ in
$G_n$ that has $2^{m}$ distinct elements up to overall phase,
there exists a set $S$ of $m$ independent generators for $V$ of the forms
$\{\bar{Z}_1,\bar{Z}_2,\cdots,\bar{Z}_\ell,\bar{X}_1,\cdots,\bar{X}_{m-\ell}\}$
where $m/2 \leq \ell \leq m$, such that $[ \bar{Z}_i , \bar{Z}_j] =
[ \bar{X}_i , \bar{X}_j] = 0$, for all $i,j$; $[ \bar{Z}_i ,
\bar{X}_j] = 0$, for all $i \neq j$; and $\{ \bar{Z}_i , \bar{X}_i
\}= 0$,  for all $i$. Let
$V_I = \langle\bar{Z}_{m-\ell+1},\cdots,\bar{Z}_\ell\rangle$ denote the
isotropic subgroup generated by the set $S_I$ of commuting generators, and let
$V_S = \langle\bar{Z}_1,\cdots,\bar{Z}_{m-\ell},\bar{X}_1,\cdots,\bar{X}_{m-\ell}\rangle$
denote the symplectic subgroup generated by the set $S_E$ of
anti-commuting generator pairs. Then, with slight abuse of the
notation, $V = \langle V_I , V_S\rangle$ denotes that $V$ is generated
by subgroups $V_I$ and $V_S$.

\begin{proof}  We sketch a constructive procedure for finding the sets of
generators, which is analogous to the Gram-Schmidt procedure for finding an
orthonormal basis for a vector space.  Suppose one has any set of generators
$\{g_1, \ldots, g_m\}$ for the subgroup $\cV$.  We will successively find new
sets of generators for $\cV$ while assigning the generators we find either to
the set $S_I$ of isotropic generators or the set $S_E$ of symplectic generators.
The generators of $S_E$ come in anticommuting pairs.

We start by assigning the anticommuting pairs.  Suppose that one has so far assigned
$k$ pairs of generators to $S_E$.  Go through the list of as-yet-unassigned
generators, and find any two generators $g$ and $h$ that anticommute.  These
will be our next symplectic pair.  First, however, we must make them commute
with all the remaining generators.  Go through the whole list of unassigned
generators (except $g$ and $h$).  If a generator anticommutes with $g$,
then multiply that generator by $h$ and replace it.  The new generator will
commute with $g$, and the new set generates the same group.  Similarly, if a generator
anticommutes with $h$, multiply it by $g$ and replace it.  (Note that some
generators may be multiplied by both $g$ and $h$, and some by neither.)
Once we have gone through the entire list, we will have found a new set of
generators such that $g$ and $h$ anticommute and all other generators
commute with them both.  Relabel $g$ as $\bar{Z}_{k+1}$ and $h$ as
$\bar{X}_{k+1}$, assign the pair to $S_E$, and  let $k\rightarrow k+1$.

Continue this procedure until either all the generators have been assigned
to $S_E$, or all of the remaining unassigned generators commute.  If there
are a total of $m-\ell$ pairs of generators in $S_E$, we relabel the remaining
generators $\bar{Z}_{m-\ell+1}, \ldots, \bar{Z}_\ell$ and assign them all
to $S_I$.  This procedure is optimal, in the sense that it produces the
minimum number of symplectic pairs (and hence the minimum number
of ebits for the code).
\end{proof}

For the group $S$ that we are considering, generated by (\ref{setM}),
we can follow the steps of the procedure outlined in the lemma to
construct such a set of independent generators.  We start by taking the
first generator $M_1$ and labeling it $\bar{Z}_1$.  We then find the first
anticommuting generator---which happens to be $M_2$---and label
it $\bar{X}_1$.  We must then eliminate any anticommutation with
the remaining generators.  $\bar{Z}_1$ anticommutes with $M_3$ and
$M_4$, so we multiply each of them by $\bar{X}_1$.  $M_4$
anticommutes with $\bar{X}_1$, so we multiply it by $\bar{Z}_1$.
The two new generators that result commute with $\bar{Z}_1$ and
$\bar{X}_1$, and also commute with each other; so we label them
$\bar{Z}_2$ and $\bar{Z}_3$.  The resulting set of generators is:
\begin{equation}
\begin{array}{ccccc}
\bar{Z}_1= & Z & X & Z & I   \\
\bar{X}_1= & Z & Z & I & Z   \\
\bar{Z}_2= & Y & X & X & Z   \\
\bar{Z}_3= & Z & Y & Y & X
\end{array} .
\label{example_generators}
\end{equation}
so that $\cS_S = \langle\bar{Z}_1,\bar{X}_1\rangle$,
$\cS_I = \langle\bar{Z}_2,\bar{Z}_3\rangle$, and $\cS=\langle\cS_I,\cS_S\rangle$.

The choice of the notation $\bar{Z}_i$ and $\bar{X}_i$ is not
accidental: these generators have exactly the same commutation
relations as Pauli operators $Z_i$ and $X_i$ on a set of
qubits labeled by $i$.  We now see that the subgroup they generate
matches one-to-one with this simpler subgroup.  Let $B$ be the group
generated by the following set of Pauli operators:
\begin{equation}
\begin{array}{ccccc}
Z_1= & Z & I & I & I  \\
X_1= & X & I & I & I  \\
Z_2= & I & Z & I & I  \\
Z_3= & I & I & Z & I  \\
\end{array} .
\label{setB}
\end{equation}
From the previous lemma, $B = \langle B_I,B_S\rangle$, where
$B_S = \langle Z_1,X_1\rangle$ and $B_I = \langle Z_2,Z_3\rangle$. Therefore, groups $B$ and
$S$ are {\it isomorphic}, which is denoted $\cB\cong\cS$.  We can
relate $S$ to the group $B$ by the following lemma
\cite{BFG05}:

\medskip

\noindent {\it Lemma 2.} \,\, If $\cB$ and $\cS$ are both subgroups of $\cG_n$,
and $\cB\cong\cS$, then there exists a
unitary $U$ such that for all  $B\in\cB$  there exists an
$S\in\cS$ such that $B = U S U^{-1}$ up to an overall phase.

\begin{proof}

We only sketch the proof here.  First, apply Lemma 1 to both $B$ and
$S$ to find a set of generators for each group in the standard form.  We call
the generators of $B$ $\{\bar{Z}_1,\ldots,\bar{Z_r},\bar{X}_1,\ldots,\bar{X}_c\}$,
and the generators of $S$ $\{\hat{Z}_1,\ldots,\hat{Z_r},\hat{X}_1,\ldots,\hat{X}_c\}$,
where $r=s+c$.  (Because $\cB\cong\cS$ the parameters $r$ and $c$ must be the
same for both groups.)  The symplectic subgroup of $B$ is generated by
$\bar{Z}_1,\ldots,\bar{Z}_c,\bar{X}_1,\ldots,\bar{X}_c$, the isotropic
subgroup of $B$ is generated by $\bar{Z}_{c+1},\ldots,\bar{Z}_r$, and similarly
for $S$.

Starting from the set of generators for $B$, we add generators to get
a complete set of generators for $G_n$:  first we add symplectic partners
$\bar{X}_{c+1},\ldots,\bar{X}_r$ for the isotropic generators $\bar{Z}_{c+1},\ldots,\bar{Z}_r$,
and then we add $n-r$ additional symplectic pairs
$(\bar{Z}_{r+1},\bar{X}_{r+1}),\ldots,(\bar{Z}_n,\bar{X}_n)$ to get a complete set of
generators.  In exactly the same way, we add new generators $\hat{X}_{c+1},\ldots,\hat{X}_r$
and $(\hat{Z}_{r+1},\hat{X}_{r+1}),\ldots,(\hat{Z}_n,\hat{X}_n)$ to the generators
for $S$ to get a different complete set of generators for $G_n$.

If any of the generators have eigenvalues $\pm i$, multiply them by $i$, so that all
generators have eigenvalues $\pm1$.  (This is allowed because we are ignoring
the overall phases.)  Now define the following two states:
$\ket{\bar{\mathbf0}}$ is the simultaneous $+1$ eigenstate of the generators
$\bar{Z}_1,\ldots,\bar{Z}_n$, and
$\ket{\hat{\mathbf0}}$ is the simultaneous $+1$ eigenstate of the generators
$\hat{Z}_1,\ldots,\hat{Z}_n$. Starting from these two states, we construct
two orthonormal bases.  Let $\mathbf{b} = (b_1,\ldots,b_n)$ be a string of $n$ bits.
Define the basis states
$\ket{\bar{\mathbf{b}}} = \bar{X}_1^{b_1}\cdots\bar{X}_n^{b_n}\ket{\bar{\mathbf0}}$ and
$\ket{\hat{\mathbf{b}}} = \hat{X}_1^{b_1}\cdots\hat{X}_n^{b_n}\ket{\hat{\mathbf0}}$.
It is easy to show that these form two orthonormal bases.  Therefore we can
define a unitary operator
\[
U = \sum_{\mathbf{b}} \ket{\bar{\mathbf{b}}} \bra{\hat{\mathbf{b}}} .
\]

One can now show that $\bar{Z}_j = U \hat{Z}_j U^\dagger$ and
$\bar{X}_j = U \hat{X}_j U^\dagger$ for all $j$.  This implies that any
operator in $\cS$ can be mapped to a corresponding operator in $\cB$,
by writing the two operators as products of corresponding generators.

\end{proof}

As a consequence of this lemma, the error-correcting codes
$\cC(B)$ and $\cC(S)$ are also related by a unitary
transformation.  In what follows, we will use the straightforward
group $B$ to discuss the error-correcting conditions for an EAQECC,
and then translate the results back to the code $\cC(S)$.
As we will see, the unitary $U$ constructed in the lemma can be
thought of as the {\it encoding} operator for the code $\cC(S)$.
Note that while this unitary has been expressed in abstract terms,
there are efficient techniques to directly find a quantum circuit for
$U$ in terms of CNOTs, Hadamards, and phase gates \cite{CG97,GRB03,WBconv1,WBconv2}.

\subsection{Anticommuting pairs and entanglement}

What is the code space $\cC(B)$ described by $B$ in (\ref{setB})?
Because $B$ is not a commuting group, the usual definition of a QECC
$\cC(B)$ does not apply, since the generators do not have a common
$+1$ eigenspace.  However, by {\it extending} the generators, we can find
a new group that {\it is} Abelian, and for which the usual definition of code
space {\it does} apply.  The qubits of the codewords will be embedded in a
larger space.  Notice that we can append a $Z$ operator at the end of
$Z_1$, a $X$ operator at the end of $X_1$, and an identity at the
end of $Z_2$ and $Z_3$, to make $B$ into a new Abelian group $B_e$:
\begin{equation}
\begin{array}{ccccc|c}
Z_1'= & Z & I & I & I & Z \\
X_1'= & X & I & I & I & X \\
Z_2'= & I & Z & I & I & I \\
Z_3'= & I & I & Z & I & I
\end{array} .
\label{extendedB}
\end{equation}
The four original qubits are possessed by Alice
(the sender), but the additional qubit is possessed by Bob (the
receiver) and is not subject to errors.  Let $B_e$ be the extended
group generated by $\{Z_1',X_1',Z_2',Z_3'\}$.  We define the code
space $\cC(B)$ to be the simultaneous $+1$ eigenspace of all
elements of $B_e$, and we can write it down explicitly in this
case:
\begin{equation}
\cC(B)=\{\ket{\Phi}^{AB}\ket{0}\ket{0}\ket{\psi}\},
\end{equation}
where $\ket{\Phi}^{AB}$ is the maximally entangled state
$(\ket{00} + \ket{11})/\sqrt2$ shared between Alice and Bob,
and $\ket{\psi}$ is an arbitrary single-qubit pure state.  (Bob's qubit
corresponds to the fifth column in (\ref{extendedB}).)
Because entanglement is used, this is an EAQECC.  The
number of ebits $c$ needed for the encoding is equal to the number
of anti-commuting pairs of generators in $B_S$. The number of
ancilla bits $s$ equals the number of independent generators in
$B_I$. The number $k$ of encoded qubits equals to $n-c-s$. Therefore,
$\cC(B)$ is a $[[4,1;1]]$ EAQECC with zero net rate:  $n=4$, $c=1$,
$s=2$ and $k=1$. Note that zero net rate does not mean that no
qubits are transmitted by this code!  Rather, it implies that the
number of ebits needed is equal to the number of qubits transmitted.
In general, $k-c$ can be positive, negative, or zero.

Now we will see how the error-correcting conditions are related to the
generators of $B$.  If an error $E_a \otimes I^B$ anticommutes
with one or more of the operators in $\{Z_1',X_1',Z_2',Z_3'\}$, it can
be detected by measuring these operators.  This will only happen if
the error $E_a$ on Alice's qubits anticommutes
with one of the operators in the original set of generators
$\{Z_1,X_1,Z_2,Z_3\}$, since the entangled bit held by Bob is assumed
to be error-free. Alternatively, if  $E_a\otimes I^B \in B_e$, or
equivalently $E_a \in B_I$, then $E_a$ does not corrupt the
encoded state. (In this case we call the code {\it degenerate}.) Altogether,
$\cC(B)$ can correct a set of errors $\cE_0$ if and only if
$E_a^\dagger E_b \in B_I \cup (G_4-Z(B))$ for all
$E_a,E_b\in\cE_0$.

With this analysis of $B$, we can go back to determine the
error-correcting properties of our original stabilizer $S$.  We
can construct a QECC from a nonabelian group $S$ if entanglement
is available, just as we did for the group $B$.  We add extra
operators $Z$ and $X$ on Bob's side to make $S$ abelian as follows:
\begin{equation}
\begin{array}{ccccc|c}
\bar{Z}_1'= & Z & X & Z & I & Z  \\
\bar{X}_1'= & Z & Z & I & Z & X  \\
\bar{Z}_2'= & Y & X & X & Z & I  \\
\bar{Z}_3'= & Z & Y & Y & X & I
\end{array} .
\label{example_augmented_generators}
\end{equation}
where the extra qubit is once again assumed to be possessed by
Bob and to be error-free. Let $S_e$ be the group generated by the
above operators.  Since $B \cong S$, let $U^{-1}$ be the unitary from
Lemma 2.  Define the code space $\cC(S)$ by
$\cC(S)=U^{-1}(\cC(B))$, where the unitary $U^{-1}$ acts only
on Alice's side.  This unitary $U^{-1}$ can be interpreted as the
encoding operation of the EAQECC defined by $S$.  Observe that the
code space $\cC(S)$ is a simultaneous eigenspace of all elements
of $S_e$.  As in the case of $\cC(B)$, the code $\cC(S)$
can correct a set of errors $\cE$ if and only if
\begin{equation}
E_a^\dagger E_b \in S_I \cup (G_4-Z(S)) ,
\label{error_condition}
\end{equation}
for all $E_a,E_b\in\cE$.

The algebraic description is somewhat abstract, so let us translate
this into a physical picture.  Alice wishes to encode a single
($k=1$) qubit state $\ket\psi$ into four ($n=4$) qubits, and
transmit them through a noisy channel to Bob.  Initially, Alice and
Bob share a single ($c=1$) maximally entangled pair of qubits---one
ebit.  Alice performs the encoding operation $U^{-1}$ on her bit
$\ket\psi$, her half of the entangled pair, and two ($s=2$) ancilla
bits.  She then sends the four qubits through the channel to Bob.
Bob measures the extended generators $\bar{Z}_1', \bar{X}_1',
\bar{Z}_2'$, and $\bar{Z}_3'$ on the four received qubits plus his
half of the entangled pair.  The outcome of these four measurements
gives the error syndrome; as long as the error set satisfies the
error-correcting condition (\ref{error_condition}), Bob can correct the
error and decode the transmitted qubit $\ket\psi$.  We can see
schematically how this procedure works in Fig.~\ref{eaqecc_schematic}.

In fact, this particular example is a $[[4,1,3;1]]$ EAQECC:  it can
correct any single-qubit error.  No standard QECC with $n<5$ can
correct an arbitrary single-qubit error.  Clearly, the use of an ebit
enhances the error-correcting power of the code.  However, the need
for a shared ebit is also a limitation, since the ebit must presumably
have been shared originally by extra channel uses---this is reflected
in the fact that the net rate of the code is zero.

We have worked out the procedure for a particular example, but any
entanglement-assisted quantum error correction code will function in
the same way.  If we have $c$ anticommuting pairs of generators
$\bar{Z}_j$ and $\bar{X}_j$ in $S_E$, we resolve their anticommutation by
adding a $Z$ and $X$ operator, respectively, acting on an extra qubit
on Bob's side.  We will need one such extra qubit for each pair, or $c$
in all; these correspond to $c$ ebits initially shared between Alice
and Bob.  The particular parameters $n,k,c,s$ will vary
depending on the code.  It should be noted that the first example of
entanglement-assisted error correction produced a $[[3,1,3;2]]$
EAQECC based on the $[[5,1,3]]$ standard QECC
\cite{Bow02}.  We have now produced a completely general
description, that also eschews the need for a commuting stabilizer group.

\subsection{The symplectic representation of EAQECCs}
\label{section_symplectic_rep}

There is another very useful representation for stabilizer QECCs where tensor products of Pauli operators are represented by pairs of bit strings:  $e^{i\phi} X^{\mathbf{a}}Z^{\mathbf{b}} \rightarrow (\mathbf{a}|\mathbf{b})$, where $\mathbf{a}$ and $\mathbf{b}$ are strings of bits $\mathbf{a} \equiv a_1 a_2 \cdots a_n$ and $\mathbf{b} \equiv b_1 b_2 \cdots b_n$, and we use the power notation:
\begin{equation}
X^{\mathbf{a}} \equiv X^{a_1} \otimes X^{a_2} \otimes \cdots \otimes X^{a_n} , \ \ \ \ \ 
Z^{\mathbf{b}} \equiv Z^{b_1} \otimes Z^{b_2} \otimes \cdots \otimes Z^{b_n} .
\end{equation}
The overall phase is lost in this representation, but for our present purposes this is not important.  We also define a map in the other direction, from a pair of bit strings to a Pauli operator:  $\mathbf{w} \rightarrow N_{\mathbf{w}} \equiv X^{\mathbf{a}}Z^{\mathbf{b}}$, where $\mathbf{w} = (\mathbf{a}|\mathbf{b})$.  

This symplectic representation has many advantages.  Up to a phase, multiplication of Pauli operators is given by binary vector addition.  A set of generators $g_1, \ldots, g_{n-k}$ for the stabilizer is given by a set of $n-k$ binary strings of  length $2n$, which we write as the rows of a matrix.  We can thus represent a quantum stabilizer code by a quantum check matrix, quite analogous to a classical linear code:
\begin{equation}
g_1, \ldots, g_{n-k} \longrightarrow \hat{H} = \left( H_X | H_Z \right) ,
\end{equation}
where $H_X$ and $H_Z$ are $(n-k)\times n$ binary matrices.  The rows of $\hat{H}$ represent the stabilizer generators, and the rowspace of $\hat{H}$ represents the full set of stabilizer operators.

While this symplectic representation does not include the overall phase of Pauli group elements, it does capture the commutation relations between different elements of the Pauli group.  If two group elements $g$ and $g'$ are represented by the strings $(\mathbf{a}|\mathbf{b})$ and $(\mathbf{a}'|\mathbf{b}')$, respectively, then their commutation relation is given by the {\it symplectic inner product} of the two strings:
\begin{equation}
(\mathbf{a}|\mathbf{b}) \odot (\mathbf{a}'|\mathbf{b}') = \mathbf{a}\cdot\mathbf{b}' + \mathbf{a}'\cdot\mathbf{b} ,
\end{equation}
where $\mathbf{a}\cdot\mathbf{b}'$ denotes the usual Boolean inner product.  The symplectic inner product between two strings is either 0 or 1; if it is 0, then the operators they represent commute, otherwise they anticommute:
\[
N_{\mathbf{w}} N_{\mathbf{w}'} = (-1)^{\mathbf{w}\odot\mathbf{w}'} N_{\mathbf{w}'} N_{\mathbf{w}} .
\]
As we will see, this enables us to derive a compact formula for the amount of entanglement needed by a given code.  For a standard stabilizer QECC, the symplectic inner product between any two rows of the quantum check matrix $\hat{H}$ must be 0.

For EAQECCs, we typically use the symplectic representation in two slightly different but related ways.  First, we can represent the (in general noncommuting) generators on Alice's qubits by an $(n-k)\times{2n}$ check matrix, whose rows can have nonzero symplectic inner product.  So, for example, the generators in Eq.~(\ref{example_generators}) are represented by the following quantum check matrix:
\begin{equation}
\begin{array}{ccccc}
\bar{Z}_1= & Z & X & Z & I   \\
\bar{X}_1= & Z & Z & I & Z   \\
\bar{Z}_2= & Y & X & X & Z   \\
\bar{Z}_3= & Z & Y & Y & X
\end{array}
\longrightarrow
\left( \begin{array}{cccc|cccc}
0 & 1 & 0 & 0 & 1 & 0 & 1 & 0 \\
0 & 0 & 0 & 0 & 1 & 1 & 0 & 1 \\
1 & 1 & 1 & 0 & 1 & 0 & 0 & 1 \\
0 & 1 & 1 & 1 & 1 & 1 & 1 & 0
\end{array}\right) .
\end{equation}
It is also useful to represent the augmented operators, in which $c$ extra qubits have been added on Bob's side in order to resolve the anticommutativity of the stabilizer group.  In this case, we include the operators on Bob's side as well, with the convention that the bits representing operators on Bob's side will be listed in the $c$ rightmost columns of $H_X$ and $H_Z$.  Taking the augmented generators from (\ref{example_augmented_generators}) we get the symplectic representation
\begin{equation}
\begin{array}{ccccc|c}
\label{setS}
\bar{Z}_1'= & Z & X & Z & I & Z  \\
\bar{X}_1'= & Z & Z & I & Z & X  \\
\bar{Z}_2'= & Y & X & X & Z & I  \\
\bar{Z}_3'= & Z & Y & Y & X & I
\end{array}
\longrightarrow
\left( \begin{array}{ccccc|ccccc}
0 & 1 & 0 & 0 & 0 & 1 & 0 & 1 & 0 & 1 \\
0 & 0 & 0 & 0 & 1 & 1 & 1 & 0 & 1 & 0 \\
1 & 1 & 1 & 0 & 0 & 1 & 0 & 0 & 1 & 0 \\
0 & 1 & 1 & 1 & 0 & 1 & 1 & 1 & 0 & 0
\end{array}\right)
\end{equation}
where the fifth and tenth columns represent the $X$ and $Z$ parts, respectively, of the operators on Bob's qubit.

\subsection{The Canonical Code}

Consider the following trivial encoding operation defined by
\begin{equation}
\ket{\varphi}  \mapsto
\ket{\Phi}^{\otimes c} \ket{\bo} \ket{\varphi} .
\label{tren}
\end{equation}
In other words, registers containing $\ket{\bo}$ (of size $s = n-k-c$
qubits) and $\ket{\Phi}^{\otimes c}$ ($c$ ebits shared beween Alice
and Bob) are appended to the register containing the ``encoded''
information $\ket{\varphi}$ (of size $k$ qubits).  To make the comparison
to other codes, we can think of Alice appending the ancillas and her
halves of the entangled pairs, and then applying an encoding
unitary that is simply the identity $I$.  What errors can we
correct with such a simple-minded encoding?

\medskip

\noindent{\bf Proposition 1.} \,
The code given by (\ref{tren}) and a suitably defined decoding map $\cD_0$
can correct an error set (represented in symplectic form)
\begin{eqnarray}
\cS_0 &=& \{  \left( \ba_1, \bb_1, \bal(\ba_1, \ba_2, \bb_1)  | \ba_2, \bb_2, \bbe(\ba_1,
\ba_2, \bb_1)  \right): \nonumber\\
&& \ba_1, \ba_2 \in (\bbZ_2)^{c},\bb_1, \bb_2 \in (\bbZ_2)^{s} \},
\label{sdef}
\end{eqnarray}
for any pair of known functions $\bal,\bbe: (\bbZ_2)^{c} \times
(\bbZ_2)^{c} \times (\bbZ_2)^{s} \rightarrow (\bbZ_2)^{k}$.
To put this more intuitively:  an error set is correctable by the canonical code
if every error in the error set leaves a record in the appended registers
that determines exactly what has been done to the information qubits
and how to undo it.  (This is why the functions $\bal$ and $\bbe$ must
be known functions.)


\begin{figure}
\centering
\includegraphics[width=5in]{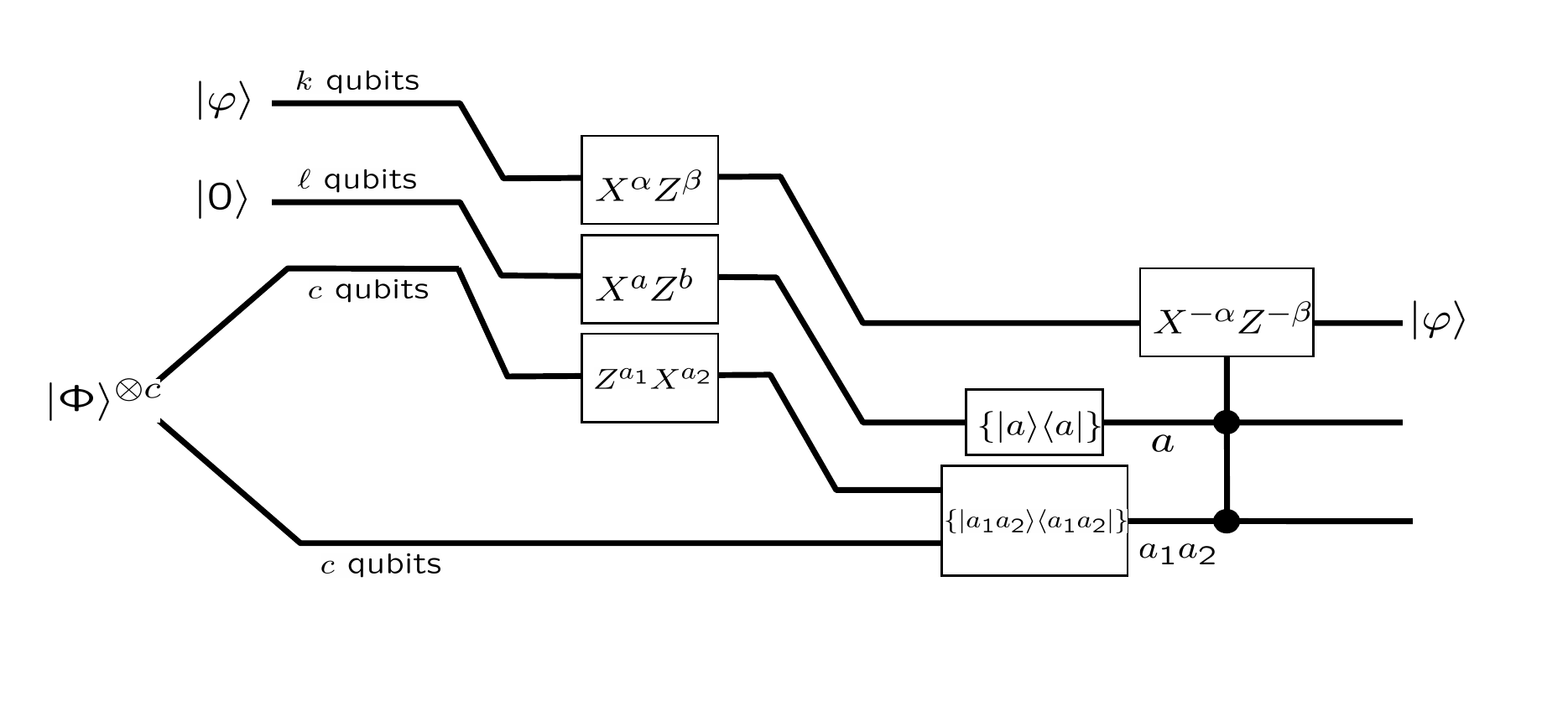}
\caption{The canonical code.  Note that this code differs from a general EAQECC only in that the encoding unitary used is the identity, and we assume a particular form for the errors.} \label{figdumb}
\end{figure}

\begin{proof}
The protocol is shown in figure \ref{figdumb}. After applying an
error $N_\bu$ with
\begin{equation}
\bu = \left( \ba_1, \bb_1, \bal(\ba_1, \ba_2, \bb_1)  | \ba_2, \bb_2, \bbe(\ba_1, \ba_2, \bb_1) \right)
\label{boo}
\end{equation}
the encoded state $\ket{\Phi}^{\otimes c} \ket{\bo} \ket{\varphi}$ becomes (up
to a phase factor)
\begin{eqnarray}
&& (X^{\ba_1} Z^{\ba_2} \otimes I) \ket{\Phi}^{\otimes c}
\otimes X^{\bb_1} Z^{\bb_2} \ket{\bo}
\otimes X^{\bal(\ba_1, \ba_2, \bb_1)} Z^{\bbe(\ba_1, \ba_2, \bb_1)} \ket{\varphi}
 \nonumber\\
&=&  \ket{\ba_1, \ba_2} \otimes \ket{\bb_1}
\otimes X^{\bal(\ba_1, \ba_2, \bb_1)} Z^{\bbe(\ba_1, \ba_2, \bb_1)} \ket{\varphi} ,
\label{zadnja}
\end{eqnarray}
where $\ket{\bb_1} =   X^{\bb_1} \ket{\bo}$ and
$\ket{\ba_1, \ba_2} = (X^{\ba_1} Z^{\ba_2} \otimes I)
\ket{\Phi}^{\otimes c}$. As the vector $(\ba_1, \ba_2, \bb_1, \bb_2)^\T$
completely specifies the error $\bu$, it is called the \emph{error
syndrome}. The state (\ref{zadnja}) only depends on the
\emph{reduced syndrome} $\br = (\ba_1, \ba_2, \bb_1)^\T$. In effect,
$\bb_1$ and $(\ba_1, \ba_2)$ have been encoded into the ancillas and ebits
using plain and superdense coding, respectively. Bob, who holds the entire state
(\ref{zadnja}) at the end, may identify the reduced syndrome by a simple
projective measurement. Bob simultaneously measures
the $Z^{\bee_1} \otimes Z^{\bee_1}, \dots, Z^{\bee_c} \otimes Z^{\bee_c}$
observables to decode $\ba_1$ (where the operators to the right of the $\otimes$
symbol act on Bob's halves of the ebits),
the $X^{\bee_1} \otimes X^{\bee_1}, \dots, X^{\bee_c} \otimes X^{\bee_c}$
observables to decode $\ba_2$,
and the $Z^{\bee_{c+1}},\dots, Z^{\bee_{c+s}}$ observables to decode $\bb_1$,
where $\{\bee_j\}$ are the standard basis vectors.
He then performs $Z^{\bbe(\ba, \ba_1, \ba_2)}
X^{\bal(\ba, \ba_1, \ba_2)}$ on the remaining $k$ qubit system,
leaving it in the original state $\ket{\varphi}$.

Since the goal is the transmission of quantum information, no actual
measurement is necessary. Instead, Bob can perform the CPTP map
$\cD_0$ consisting of the controlled  unitary
\begin{equation}
U_{0 \, \rm{dec}} = 
\sum_{\ba_1, \ba_2, \bb_1} \proj_{\ba_1,\ba_2} \otimes \proj_{\bb_1} \otimes
Z^{\bbe(\ba_1, \ba_2, \bb_1)} X^{\bal(\ba_1, \ba_2, \bb_1)} ,
\label{CanonicalUnitaryDecoder}
\end{equation}
followed by discarding the first two subsystems.  Here we use projection
operators $\proj_{\ba_1,\ba_2} = \ket{\ba_1,\ba_2}\bra{\ba_1,\ba_2}$ and
$\proj_{\bb_1} = \ket{\bb_1}\bra{\bb_1}$.  (Note that equations very similar to (\ref{boo}--\ref{CanonicalUnitaryDecoder}) arise in the theory of continuous variable error-correcting codes, as described in \cite[Chapter 22]{LB13}.
\end{proof}
\medskip

The above code is  \emph{degenerate} with respect to the error set
$S$, which means that the error can be corrected without knowing the
full error syndrome.  This is because the $Z^\bb$ operator acting on
the ancillas has no effect beyond a trivial global phase.  It is our assumption
that the errors acting on the information qubits did not depend on $\bb$
that makes this error set correctable.  This kind of restriction will always
be necessary---no error-correcting code can correct every possible error.
Part of the art of error correction is to choose a code whose correctable
error set matches, as closely as possible, the actual physical errors that
are likely to happen.

We can characterize our code in terms of the check matrix $\H$ given by
\begin{equation}
\label{uno1}
\H = \left(\begin{array}{c}
\H_I \\
\H_S
\end{array} \right),
\end{equation}
\begin{equation}
\label{uno2}
\H_I = \left(\begin{array}{ccc|ccc} \bo_{s \times c} & \bo_{s \times s} & \bo_{s \times k} & \bo_{s
\times c} & \bI_{s \times s} & \bo_{s \times k}
\end{array} \right),
\end{equation}
\begin{equation}
\label{uno3}
\H_S = \left(\begin{array}{ccc|ccc}
 \bI_{c \times c} & \bo_{c \times s} & \bo_{c \times k}  &
 \bo_{c \times c} & \bo_{c \times s} & \bo_{c \times k}  \\
\bo_{c \times c} & \bo_{c \times s} & \bo_{c \times k}
 & \bI_{c \times c} & \bo_{c \times s} & \bo_{c \times k}
\end{array} \right),
\end{equation}
with $s = n-k-c$.

The row space of $\H$ decomposes into a direct sum of the
{\it isotropic} subspace, given by the row space of $\H_I$ and the {\it symplectic}
subspace, given by the row space of $\H_S$.
Define the \emph{symplectic code} corresponding to $\H$ by
$$
\cC_0 = {\rm rowspace}(\H)^\perp
$$
where
$$
V^\perp = \{\bw: \bw \odot \bu^\T = 0, \,\, \forall \bu \in V\}.
$$
Note that $(V^\perp)^\perp = V$. 
Then $C_0^\perp = {\rm rowspace}(\H)$, ${\rm iso}(C_0^\perp) = {\rm rowspace}(\H_I)$ and
 ${\rm symp}(C_0^\perp) = {\rm rowspace}(\H_S)$.

Just as with standard stabilizer codes (as described earlier in this chapter),
each row of the check matrix $\H$ maps to a stabilizer generator, and the complete set
of vectors in the row space of $\H$ maps onto the complete set of stabilizer operators
(up to phases).  The rows of $\H_S$ come in nonorthogonal pairs under the symplectic
inner product; these nonorthogonal pairs of rows map
into anticommuting pairs of generators in the stabilizer.  Each such pair requires one extra
qubit on Bob's side to resolve the noncommutativity of the stabilizer, and hence
requires one shared ebit between Alice and Bob at the encoding step.  The number of
ebits used in the code is therefore
$$
c = \frac{1}{2}\dim {\rm rowspace}(\H_S)
$$
and the number of encoded qubits is
\begin{eqnarray}
k &=& n - \dim {\rm rowspace}(\H_I) - {\small \frac{1}{2}} \dim {\rm rowspace}(\H_S) \nonumber\\
&=&  n - \dim {\rm rowspace}(\H) + c.
\end{eqnarray}
The code parameter $\hat{k}: = k - c$, which is the
number of encoded qubits minus the number of ebits used, is
independent of the symplectic structure of $\H$:
\begin{equation}
\hat{k} = n - \dim {\rm rowspace}(\H).
\end{equation}
We see that the quantity $k/n$ is the rate of the code, and $\hat{k}/n$ is the net rate.

The correctable error set $S_0$ of this symplectic code can be described in terms of $\H$:

\noindent{\bf Proposition 2.} \,
The set $\cS_0$ of errors correctable by the canonical code (in symplectic form)
is such that, if $\bu, \bu' \in \cS_0$ and $\bu \neq \bu'$, then either
\begin{enumerate}
\item $\bu - \bu' \not\in \cC_0$ (equivalently:
 $\H \odot (\bu - \bu')^\T \neq \bo^\T$), or
\item $\bu - \bu' \in {\rm iso}(\cC_0^\perp)$
(equivalently: $\bu - \bu' \in {\rm rowspace}(\H_I)$).
\end{enumerate}

\begin{proof}
If $\bu$ is given by (\ref{boo}) then $\H \odot \bu^\T = \br = (\ba_1,
\ba_2, \bb_1)^\T$, the reduced error syndrome. By definition
(\ref{sdef}), two distinct elements of $S_0$ either have different
reduced syndromes $(\ba_1, \ba_2, \bb_1)$ (condition 1) or they differ
by a vector of the form $( \bo, \bo, \bo |  \bo,  \bb,  \bo)$
(condition 2).  In the first case, the two errors can be distinguished,
and appropriate different corrections applied.  In the second
case, the two errors have the same reduced syndrome and
hence cannot be distinguished, but they are both correctable
by the same operation.  Observe that condition 1 is analogous to
the usual error correcting condition for classical codes \cite{FJM77},
while condition 2 is an example of the quantum phenomenon of
degeneracy.
\end{proof}

\medskip

The parity check matrix $\H$ also specifies the encoding and
decoding operations. The space ${\cH_2}^{\otimes k}$ is encoded into the
\emph{codespace} defined by
\[
\cC_0 = \{   \ket{\Phi}^{\otimes c} \ket{\bo} \ket{\varphi}  : \ket{\varphi}
\in {\cH_2}^{\otimes k}\}.
\]
It is not hard to see that the codespace is the simultaneous $+1$
eigenspace of the commuting operators:
\begin{enumerate}
\item $I \otimes Z^{\bee_i} \otimes I \otimes I, \,\,i= 1, \dots, s$;
\item $Z^{\bee_j} \otimes I \otimes I \otimes Z^{\bee_j}, \,\, j= 1, \dots, c$;
\item $X^{\bee_j} \otimes I \otimes I \otimes X^{\bee_j} , \,\, j= 1, \dots, c$.
\end{enumerate}
Above, in operators written $A_1\otimes A_2 \otimes A_3 \otimes B$ the first three operators $A_1$, $A_2$, $A_3$ act on Alice's qubits, and the fourth operator $B$ on Bob's. Define the check matrix
\begin{equation}
H_B =
 \left(\begin{array}{c|c}
 \bo_{s \times c}  & \bo_{s \times c} \\
 \bI_{c \times c}  & \bo_{c \times c} \\
 \bo_{c \times c}  & \bI_{c \times c}
\end{array} \right).
\label{eq:beba}
\end{equation}
Define the \emph{augmented} parity check matrix $\H_{\rm aug} = (\H, H_B)$
\begin{equation}
\H_{\rm aug} =
\left(\begin{array}{cccc|cccc} \bo_{s \times c} & \bo_{s
\times s} & \bo_{s \times k} & \bo_{s \times c} & \bo_{s
\times c} & \bI_{s \times s} &
\bo_{s \times k} & \bo_{s \times c} \\
\bI_{c \times c} & \bo_{c \times s} & \bo_{c \times k} & \bI_{c
\times c} & \bo_{c \times c} & \bo_{c \times s} & \bo_{c \times k} & \bo_{c \times c} \\
\bo_{c \times c} & \bo_{c \times s} & \bo_{c \times k} &
\bo_{c\times c} & \bI_{c \times c} & \bo_{c \times s} & \bo_{c
\times k} & \bI_{c \times c}
\end{array} \right).
\end{equation}
Observe that ${\rm rowspace}(\H_{\rm aug})$ is purely isotropic. The
codespace is now described as the simultaneous $+1$ eigenspace of
\[
\{  N_\bw: \bw \in {\rm rowspace}(\H_{\rm aug})\},
\]
or, equivalently that of
\[
G_0 =  \langle  N_\bw: \bw {\rm \,\,is \,\, a \,\, row \,\, of \,\,} \H_{\rm aug} \rangle.
\]
The decoding operation $\cD_0$ is also described in terms of $\H$.
The reduced syndrome $\br = \H \odot \bu^\T$ is obtained by
simultaneously measuring the observables in $\cG_0$. The reduced
error syndrome corresponds to an equivalence class of possible errors
$\bu \in S_0$ that all have an identical effect on the codespace, and can
all be corrected by the same operation.  Bob performs
$\hat{N}_{\bu} = \hat{N}_{-\bu}$ to undo the error.



\subsection{General codes and code parameters}

The canonical code described above is not very useful for actual error correction.  It corrects a set of highly nonlocal errors, of a type that are not likely to occur in reality.  However, it is very helpful in clarifying how the process of error correction works, and the role of the resources used in encoding.  Each added ancilla can contain, in principle, one bit of information about any errors that occur.  Each shared ebit between the sender and receiver can contain two bits of information about errors.  This information about the errors is retrieved by measuring the stabilizer generators.  After Alice's qubits have passed through the channel, Bob has all the qubits and can perform these measurements.  

A general EAQECC has exactly the same algebraic structure as the canonical code.  The encoded state, before going through the channel, has $c$ ebits of entanglement with the $c$ qubits on Bob's side.  Corresponding to this entanglement are $c$ pairs of stabilizer generators.  Each pair of generators has the following structure:  a pair of anticommuting operators on Alice's qubits, tensored with a $Z$ and an $X$, respectively, on one of the qubits on Bob's side to ``resolve'' the anticommutativity.    These pairs of generators are the symplectic pairs of the EAQECC.  In addition to these pairs, there can be single generators that act as the identity on all of Bob's qubits.  These are the isotropic generators.  For an $[[n,k,d;c]]$ EAQECC, each of the stabilizer elements acts on $n+c$ qubits, $n$ on Alice's side and $c$ on Bob's side;  there is a set of $n-k$ generators for the stabilizer chosen in a standard form, with $2c$ symplectic pairs of generators and $s=n-k-2c$ isotropic generators.

By Lemma 2, if two sets of Pauli operators can be put in one-to-one correspondence so that they have the same commutation relations, there is a unitary transformation that turns one set into the other (up to a phase).  The $n-k$ generators of an $[[n,k;c]]$ EAQECC can be mapped by a unitary operator $U_E$ to the generators of a canonical code.

It is important to realize is that the canonical code describes the {\it input} of the EAQECC {\it before encoding}.  It describes a set of $k$ information qubits, $c$ ebits, and $s$ ancillas.  The unitary $U_E$ transforms the stabilizer for this input into the stabilizer for the EAQECC.  Therefore  $U_E$ is a unitary encoding operator for this code.  It transforms the localized states of the input qubits into nonlocal, entangled states spread over all the qubits.  By this transformation, the correctable error set is transformed from the strange, nonlocal errors of the canonical code to a more physically realistic set of errors (most commonly, localized errors on the individual qubits).

\section{Constructing EAQECCs from classical linear codes}

\subsection{Mapping $GF(4)$ to Pauli Operators}
\label{eaqecc_classical_construction}

We will now examine the $[[4,1;1]]$ EAQECC used as an example
in Section~\ref{constructingEAQECCs} above, and show
that it can be derived from a classical non-dual-containing
quaternary $[4,2]$ code. This is a generalization of the well-known
CRSS construction for standard QECCs \cite{CRSS98}.

First, note that this $[[4,1;1]]$ code is non-degenerate, and can
correct an arbitrary one-qubit error. (The distance $d$ of
the code $\cC(S)$ is 3.) This is because the $12$ errors $X_i,
Y_i$ and $Z_i$, $i = 1, \dots, 4$, have distinct non-zero error
syndromes. $X_i$ denotes the bit flip error on $i$-th qubit, $Z_i$
denotes the phase error on $i$-th qubit, and $Y_i$ means that both a
bit flip and phase flip error occur on the $i$-th qubit.  It
suffices to consider only these three standard one-qubit errors,
since any other one-qubit error can be written as a linear
combination of these three errors and the identity.

Next, we define the following map between the Pauli operators, symplectic
strings, and elements of $GF(4)$, the field with 4 elements:
\[
\begin{array}{|c||c|c|c|c|} \hline
G &  I & X & Y & Z \\  \hline
{\rm Symplectic} & 00 & 10 & 11 & 01 \\ \hline
GF(4) &  0 & \overline{\omega} & 1 & \omega \\ \hline
\end{array}
\]
Addition of symplectic strings is by bitwise XOR.  The elements of $GF(4)$ obey a simple addition rule:  $0+x=x+0=x$ and $x+x=0$ for all $x\in GF(4)$, $\overline{\omega}+1 = \omega$, $1+\omega=\overline{\omega}$, and $\omega+\overline{\omega}=1$, where addition is commutative.  Multiplication is also commutative:  $0\cdot x = 0$ and $1\cdot x = x$ for all $x\in GF(4)$, $\omega\cdot\omega = \overline{\omega}$, $\overline{\omega}\cdot\overline{\omega} = \omega$, and $\omega\cdot\overline{\omega} = 1$.  Note that under this map, addition in $GF(4)$ (or of symplectic strings) corresponds to multiplication of the Pauli operators, up to an overall phase.  So multiplication of two elements of $G_n$ corresponds to addition of two $n$-vectors over $GF(4)$, up to an overall phase.  The same correspondence applies between multiplication of the Pauli operators and bitwise addition (or XOR) of the symplectic strings.

We can define a linear code over $GF(4)$ in exactly the same way we do for a binary or symplectic code, by writing down a check matrix.  Consider the matrix
\begin{equation}
H_4 = \left(\begin{array}{cccc}
1 & \omega  & 1 & 0  \\
1 & 1  & 0 & 1
\end{array} \right).
\end{equation}
The code is the null space of $H_4$ over $GF(4)$---the set of vectors orthogonal to all the rows of the check matrix.  $H_4$ is the check matrix of a classical $[4,2,3]$ quaternary code
whose rows are not orthogonal, and 3 is the minimum distance between codewords.

From this starting point we can define a new matrix
\begin{equation}
\label{CtoQ}
\tilde{H}_4=\left(\begin{array}{c} \omega H_4 \\
\bar{\omega}H_4 \end{array}\right) =
\left(\begin{array}{cccc}
\omega & \bar{\omega}  & \omega & 0  \\
\omega & \omega  & 0 & \omega  \\
\bar{\omega} & 1 & \bar{w} & 0  \\
\bar{\omega} & \bar{\omega} & 0 & \bar{\omega}
\end{array} \right).
\end{equation}
As a quaternary code, this defines exactly the same code space as the original $H_4$, since the last two rows are multiples of the first two rows.  However, we now apply our map from $GF(4)$ to $G_4$, where each row of $\tilde{H}_4$ is mapped to a Pauli operator in $G_4$.  The resulting set of operators is exactly the set of generators $\{M_i\}$ given in (\ref{setM}).  If we map the matrix $\tilde{H}_4$ to a symplectic matrix, using the correspondence given above, we see that we get exactly the same symplectic representation for the code as in section \ref{section_symplectic_rep}.  We can carry out this procedure with any linear quaternary code.

The generators corresponding to the rows of the matrix $\tilde{H}_4$ will not
commute, in general.  In that case, we must apply Lemma 1 to find a
new set of generators comprising isotropic generators (that commute with
all other generators) and symplectic pairs of generators (that anticommute
with each other but commute with all other generators).  Extra operators can
be added on Bob's side to resolve the anticommutativity, as described
above.  This produces an EAQECC using $c$ ebits, where $c$ is the
number of symplectic pairs.

In the example above, we get a $[[4,1,3;1]]$ EAQECC from a classical $[4,2,3]$ quaternary
code. This outperforms the best 4-bit self-dual QECC currently
known, which is $[[4,0,2]]$, both in rate and distance \cite{CRSS98}.  This connection between
EAQECCs and quaternary classical codes is quite general \cite{BDH06b}.
Given an arbitrary classical $[n,k,d]$ quaternary code, we can use
(\ref{CtoQ}) to construct a non-degenerate $[[n,2k-n+c,d;c]]$ EAQECC.
The rate becomes $(2k-n)/n$ because the $n-k$ classical parity checks
give rise to $2(n-k)$ quantum stabilizer generators.  This gives us a
tremendous advantage in constructing quantum codes, because the
parameters of the quantum code depend exactly on those for the classical
code:  the minimum distance is the same, and high-rate classical codes
produce high-rate (though not as high) quantum codes.

The one parameter that is not simply determined by the parameters of
the classical code is $c$, the number of ebits used.  For a standard QECC,
which uses no entanglement and hence has $c=0$, the rows of the
classical check matrix are all orthogonal.  Intuitively, we might expect
the amount of entanglement needed to grow the further the matrix is from
being self-orthogonal.  In fact, for a CRSS code of the type described above,
the amount of entanglement needed is
\begin{equation}
c = {\rm rank}\left( H_4 H_4^\dagger \right) ,
\end{equation}
where the conjugate $H_4^\dagger$ is defined by taking the transpose
of $H_4$ and interchanging $\omega \leftrightarrow \bar{\omega}$ \cite{WB08,W09}.  For
a standard QECC, in which the rows are all orthogonal, $H_4 H_4^\dagger$
will be the zero matrix.  Since this matrix is $(n-k)\times(n-k)$, the maximum
rank (and hence the maximum possible entanglement used by the code) is
$n-k$.  Such a quantum code will transmit $2k-n+c = k$ qubits---the same
as the original classical code.

A special case of this construction is where the original classical code is
{\it binary}---that is, when the check matrix $H$ uses only 0s and 1s, and
has no elements $\omega$ or $\bar{\omega}$.  In that case, the procedure
for constructing a quantum code is to create a symplectic check matrix of the form
\[
\left(\begin{array}{c | c} H & 0 \\ 0 & H \end{array}\right).
\]
The stabilizer generators corresponding to the rows of this matrix either contain
only $Z$s and $I$s or only $X$s and $I$s, and the two sets of generators are
identical in form.  In a sense, such a code can be thought of as two classical
codes superimposed, one to correct bit flips ($X$ errors) and the other to correct
phase flips ($Z$ errors).  This is the CSS construction, analogous to the way it is
defined in \cite[Chapter 2]{LB13} for standard QECCs.

Somewhat more generally, we could define a QECC using
two different classical binary codes with check matrices $H_1$ and $H_2$:
\[
\left(\begin{array}{c | c} H_1 & 0 \\ 0 & H_2 \end{array}\right).
\]
Once again the generators would use either $X$s or $Z$s, and not both; but
the generators will no longer share the same structure.  This approach may be
useful when the noise is {\it asymmetric}; for instance, dominated by phase
errors with occasional rare bit flips.  The simpler CSS code is obviously the special
case where $H_1 = H_2 = H$.  It is not necessary that the check matrices
$H_1$ and $H_2$ be the same size; $H_1$ can be $(n-k_1)\times n$ and
$H_2$ can be $(n-k_2)\times n$.  Together they produce a quantum code
that encodes $k_1 + k_2 - n + c$ information qubits in $n$ physical qubits
with the use of $c$ ebits.  For a CSS code of this type the
entanglement needed \cite{BDH06b} is
\begin{equation}
c = {\rm rank}\left( H_1 H_2^T \right) .
\end{equation}

\subsection{Performance}

\begin{table}
\centering
\begin{tabular}{|c||c|c|c|c|c|c|c|c|c|c|c|}
  \hline
  $n \backslash\hat{k}$ & 0           & 1            & 2            & 3            & 4             & 5           & 6           & 7    & 8   & 9    & 10      \\ \hline
  3              & $2$       & $\,\,\,2^*$     & $1$    & $1$     &               &             &             &       & & &     \\ \hline
  4              & $\,\,\,3^*$       & $2$    & $2$     & $1$    & $1$      &             &             &        & & &      \\ \hline
  5              & $3$   & $3$     & $2$     & $\,\,\,2^*$        & $1$     & $1$    &             &        & & &      \\ \hline
  6              & $4$       & $3$      & $2$     & $2$     & $2$         & $1$   & $1$    &         & & &     \\ \hline
  7              & $3$       & $3$     & $2$    & $2$     & $2$      & $\,\,\,2^*$       & $1$   & $1$   &  & &  \\ \hline
  8              & $4$       & $3$    & $3$     & $3$        & $2$      & $2$    & $2$       & $1$   &   $1$ &  &  \\ \hline
  9              & $4$   & $\,\,\,4^*$     & $3$    & $3$     & $2$     & $2$    & $2$    & $\,\,\,2^*$     &  $1$   &  $1$ &   \\ \hline
  10             & $\,\,\,5^*$       & $4$    & $4$     & $3$    & $3$      & $2$   & $2$    & $2$   & $2$   & $1$   &  $1$  \\ \hline

  \hline
\end{tabular}
\caption{Table of codes up to length 10.  The best minimum distance is given for a code with length $n$ and net transmission $\hat{k} = k - c$, where $k$ is the number of information qubits and $c$ the number of ebits used.  See \cite{BDH06b}.}
\label{eaqecc_table}
\end{table}

In \cite{CRSS98} a table of the best known QECCs was given. In table \ref{eaqecc_table} we
show an updated table which includes EAQEC codes.
The entries with an asterisk mark improvements over the table
from \cite{CRSS98}. All these codes were constructed from classical
quaternary codes by the construction in section \ref{eaqecc_classical_construction},
and were found by numerical search.
The corresponding classical quaternary code is available online at:

{\tt http://www.win.tue.nl/$\sim$aeb/voorlincod.html}.

\section{Catalytic QECCs}

EAQECCs make use of an additional resource---prior shared entanglement---to boost the power of an error-correcting code.  This paradigm raises several theoretical and practical questions, which address how these codes might be useful in practice, and how they fit into different families of codes that draw on different resources.

The first practical question is:  how was this entanglement shared in the first place?  If the sender and receiver communicate by a noisy quantum channel, then the entanglement would have to be shared through this channel.  Alice locally prepares EPR pairs, and sends half of each pair through the channel, either protected by a standard QECC, or making use of entanglement purification techniques \cite{BBPSSW96,BDSW96} to produce a smaller number of perfect EPR pairs.  So we see that in this context the enhanced rate of an EAQECC must be paid for by additional channel communication to establish the entanglement in the first place.  This means that for many purposes it is the {\it net rate} $(k-c)/n$ of the EAQECC that is the appropriate figure of merit.  This net rate need not be better than that achieved by a standard QECC; indeed, it is quite possible for the net rate to be zero or negative.

Even a code with a negative net rate may actually be useful in practice.  One great advantage of shared entanglement as a resource is that it is independent of the message being sent, and can in principle be prepared well ahead of time.  One natural application of EAQECCs would therefore be to a quantum network where usage varies at different times.  During periods of low usage, shared entanglement could be built up between the sender and receiver.  During periods of high usage, this shared entanglement could be drawn on to increase the rate of transmission without sacrificing error-correcting power.

EAQECCs with positive net rates can be used in other ways to improve the power and flexibility of quantum communications.  To illustrate this, let us consider the idea of a {\it catalytic} code.  Here, in addition to $n$ uses of a noisy channel, the sender may send $c$ qubits noise-free.  One obvious approach is to send $c$ information qubits using the noise-free bits, and then send $k$ more qubits through the channel by means of an $[[n,k]]$ standard QECC.  However, another way of proceeding would be to use the $c$ noise-free bits to establish $c$ ebits between Alice and Bob, and then use an $[[n,k';c]]$ EAQECC.  So long as $k' \ge k+c$ (and both codes are assumed to correct the errors in the channel), one does as well or better with the second approach as with the straightforward one.  In this case, the noise-free qubits are used to boost the rate of the noisy qubits.  We call a procedure of this type a {\it catalytic code}---the $c$ noise-free qubits enhance transmission by an amount equal to at least $c$.

In practice, noise-free transmission is not really possible.  However, it is possible to simulate it by using a standard QECC to protect the $c$ noise-free qubits.  Suppose that we start with an $[[n,k;c]]$ EAQECC.  To establish $c$ ebits with Bob, Alice would need to send $c$ qubits through the channel.  Suppose that she protects them with an $[[n',c]]$ standard QECC.  Then we can think of the full block of $n+n'$ qubits that go through the channel as a single codeword; by putting these two codes together we have constructed an $[[n+n',k]]$ standard QECC.

Of course, the second code could also be an EAQECC.  If we have an $[[n,k;c]]$ EAQECC and an $[[n',k';c']]$ EAQECC, we can send Bob's halves of the $c$ ebits in the second code block, so long as $k'\ge c$.  By putting these two code blocks together, we have constructed an $[[n+n',k+k'-c;c']]$ EAQECC.

Nor need we combine only two codes in this way.  One rather simple construction is as follows:  take an $[[n,k;c]]$ EAQECC with positive net rate, and encode Bob's halves of the $c$ ebits in another copy of the same code.  Repeat this $M$ times, and one quickly builds up a much larger block code:  an $[[Mn,M(k-c)+c;c]]$ EAQECC.  This procedure is called {\it bootstrapping}.  Two observations of this new code are quickly evident:  first, as $M$ becomes large, the rate of the code approaches the net rate $(k-c)/n$ of the original code.  Second, as $M$ becomes large, the rate of entanglement usage $c/Mn$ becomes small.  By bootstrapping we have built a large code whose rate is the net rate of the original code, and which uses relatively little entanglement.

Practically speaking there is a limit to how big $M$ can be; if the block size becomes too large, the probability that one of the sub-blocks of $n$ qubits will have an uncorrectable error becomes high, and the code will no longer be useful.  At this point, little is known about the practical performance of catalytic codes---more research is needed.

By this approach we see another use of EAQECCs:  rather than being used themselves directly, they can be used as building blocks for the construction of standard codes.  Because EAQECCs can be constructed from classical codes without the need to satisfy the dual-containing constraint, they in some cases will outperform standard codes with otherwise similar parameters.  Combining EAQECCs with each other, or with standard codes, may allow the relatively easy construction of standard codes with good error-correcting properties.  Moreover, this may lead to applications of EAQECCs beyond the quantum communication paradigm, using them to construct standard codes that might be useful in other applications (such as quantum computing).  However, at present this is pure speculation:  such applications have yet to be studied.

\section{Conclusions}

EAQECCs expand the usual paradigm of quantum error correction by allowing the sender and receiver to make use of pre-shared entanglement.  This entanglement can increase either the rate of communication or the number of correctable errors.  While entanglement can be used to improve quantum communication in other ways---for example, by sending classical information through the channel and teleporting extra qubits---in many cases EAQECCs have better performance \cite{HW10b, HW10}.

In this chapter we have presented an extension of the usual stabilizer formalism for QECCs to include entanglement-assistance.  This extension includes the usual theory of stabilizers as a special case, and has a number of advantages; in particular, it allows one to construct quantum codes from classical linear codes without having to impose the dual-containing constraint of standard codes, while retaining the usual relation between the rate and distance of the classical and quantum codes.  Moreover, if entanglement-assistance is allowed, one can in some cases outdo the performance of the best standard QECC by optimizing without the dual-containing constraint.

In quantum communication protocols involving  entanglement, the applications of EAQECCs are obvious.  However, the construction may also be useful in other cases, by using EAQECCs as {\it catalytic} codes.  In this case, we are essentially using EAQECCs as building blocks for standard codes, or for codes that use only a small amount of pre-existing entanglement.  In these constructions the net rate of the EAQECCs is the most important parameter; this net rate can be directly calculated from the classical linear code, for codes constructed from classical precursors.

Because these constructions do not require dual-containing codes, it is possible to directly derive quantum versions of highly-efficient modern codes, such as Turbo codes and LDPC codes \cite{HYH10, WHB14}.  Standard quantum versions of these codes have a number of difficulties, since the dual-containing constraint generally implies that such codes have problems in using iterative decoding algorithms.  Some examples have been studied, comparing specific EAQECCs to standard QECCs, where the EAQECCs show very good decoding behavior \cite{HBD09,HYH10}.  Much more work remains to be done here, however.

This chapter has considered only block codes.  Some work has already been done on entanglement-assisted quantum convolutional codes as well \cite{WBconv1,WBconv2}---standard convolutional QECCs are described in \cite[Chapter 9]{LB13}.  Many questions remain open; in particular, the minimum entanglement needed to produce a quantum version of a classical convolutional code is unknown, though some answers have been conjectured \cite{WB08}.  The field of entanglement-assisted codes is relatively new, and active research is ongoing to find more applications of these codes.

\bibliographystyle{alphaurl}
\bibliography{eaqecc_refs}

\end{document}